\documentclass[pra,aps,showpacs,superscriptaddress]{revtex4}
\usepackage{graphicx}
\usepackage{amsmath}
\usepackage{epsfig}
\usepackage{amssymb}
\usepackage{epstopdf}
\begin{document}
\title{Intracavity weak nonlinear phase shifts with single photon driving. }

\author{W.J. Munro}
\address{Hewlett-Packard Laboratories, Filton Road, Stoke Gifford,
Bristol BS34 8QZ, United Kingdom}
\author{K. Nemoto}
\address{National Institute of Informatics, 2-1-2 Hitotsubashi,
Chiyoda-ku, Tokyo 101-8430, Japan}
\author{G. J. Milburn}
\address{Centre for Quantum Computer Technology, University of
Queensland, St Lucia, Queensland 4072, Australia}

\begin{abstract}
We investigate  a doubly resonant optical cavity containing a Kerr nonlinear medium that couples two modes by a cross phase modulation.  One of these modes is driven by a single photon pulsed field, and the other mode is driven by a coherent state. We find an intrinsic phase noise mechanism for the cross phase shift on the coherent beam which can be attributed to the random emission times of single photons from the cavity. An application to a weak nonlinearity phase gate is discussed. 
\end{abstract}

\maketitle

\section{Introduction}
The field of quantum optics, now roughly four decades old, arose due to theoretical and technological innovations driven by the invention of the laser. From the very beginning single photon states were a central concern, especially in the context of photon antibunching. However it is probably fair to say that for most of those four decades the field has been largely concerned with Gaussian states, beginning with Glauber's work on coherence\cite{Glauber}, and subsequent work on squeezed states in the 1980s and 1990s.  Much of the formalism of quantum optics, especially quasi-probability function methods, is based on systems driven by gaussian states. In the last decade however this has changed and there is now an emphasis on one and two photon sources\cite{SPS-issue}, and their applications in  quantum information processing\cite{KLM,Kok}. The study of single photon sources driving polarisable media is just beginning\cite{Lloyd, Razavi, Peng, Lin, Schuck}.

\begin{figure}[!hbt]
\begin{center}
\includegraphics[scale=1.0]{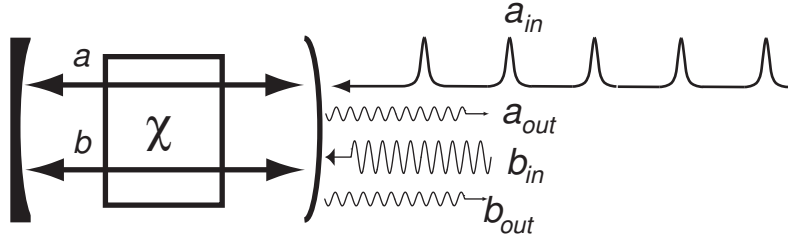}
\end{center}
\caption{An optical cavity contains a Kerr nonlinear medium and is doubly resonant for two modes, labeled a,b. Mode-a is driven by a sequence of single photon pulses, while mode-b is driven by a coherent field. }
\label{fig-1}
\end{figure}

We consider a Kerr nonlinear medium,  inside an optical cavity, that couples two-modes by cross phase modulation, see figure \ref{fig-1}.  One of these modes is driven by a pulsed field prepared in a photon number eigenstate, that is to say, each pulse has one and only one photon. The other mode is driven by a coherent state, which may be pulsed or a continuous wave. The objective is to compute the phase shift of the coherent field emitted from the cavity that is induced by the single photon. Schemes such as this form the basis of quantum gates based on weak nonlinearities\cite{Munro}. We also find an intrinsic phase noise, similar to that of \cite{Shapiro} which can be attributed to the random emission times of single photons from the cavity. 

 The cavity is doubly resonant for two modes, labeled $a,b$.  The input to mode-$a$ is a pulsed field at carrier frequency $\Omega_a$ with a single photon excitation per pulse. The input to mode-$b$ is a coherent field, either continuous or pulsed, at carrier frequency $\Omega_b$.    We first describe the single photon input states at the carrier frequency $\Omega_a$. The cavity geometry ensures mode-$a$ and mode-$b$ are coupled to quasi-one dimensional modes outside the cavity. We assume all other modes are in the vacuum state and ignore them.  In the continuum limit we define the positive frequency operator for the multi-mode field input to the cavity mode-$a$, at the input mirror, as
\begin{equation}
a_i(t)=e^{-i\Omega_a t}\frac{1}{\sqrt{2\pi}}\int_{-\infty}^\infty d\omega a_i(\omega) e^{-i\omega t}
\label{field-op}
\end{equation}
where 
\begin{equation}
[a_i(\omega_1),a_i^\dagger(\omega_2)]=\delta(\omega_1-\omega_2)
\end{equation}
Only some finite bandwidth, $B$ of these  modes are excited around the carrier frequency $\Omega_a$. At optical frequencies, $\Omega_a>>B$ which enables us to set the lower limit of integration to minus infinity in Eq.(\ref{field-op}). In this form the moment $n(t)=\langle a_i^\dagger(t)a_i(t)\rangle$ has units of ${\rm s}^{-1}$. This moment determines the 
probability per unit time (the count rate) to count a photon at the input mirror between time $t$ and $t+dt$\cite{Walls-Milb}.  Likewise, we define the positive frequency input operators for mode-$b$:
\begin{equation}
b_i(t)=e^{-i\Omega_b t}\frac{1}{\sqrt{2\pi}}\int_{-\infty}^\infty d\omega b_i(\omega) e^{-i\omega t}
\end{equation}

The multimode single  photon state is defined by\cite{Milburn-SPS}
\begin{equation}
|1\rangle=\int_{-\infty}^\infty\ d\omega\ \nu(\omega)a_i^\dagger(\omega)|0\rangle
\label{single-photon}
\end{equation}
Normalisation requires that
$
\int_{-\infty}^\infty d\omega |\nu(\omega)|^2=1
$
This state has zero average field amplitude but $
n(t)=|\nu(t)|^2
$
 where $\nu(t)$ is the Fourier transform of $\nu(\omega)$.  For this state the function $\nu(t)$ is periodic in the phase $\omega t$ and it is not difficult to choose a form with a well defined pulse sequence. However care should be exercised in interpreting these pulses. They do not represent a sequence of pulses each with one photon, rather they represent a single photon coherently superposed over all pulses. Once a photon is counted in a particular pulse, the field is returned to the vacuum state. It would be more realistic to consider a field with one and only one photon per pulse, but for this paper the simpler version will suffice.  A single photon state is a highly non-classical state with applications to quantum information processing. A review of current efforts to produce such a state may be found in \cite{SPS-issue}.

 \section{Quantum stochastic differential equations.}
 \label{SDE}

We assume that the input field for mode-$b$ is a coherent state  with a carrier frequency equal to $\Omega_b$ and an amplitude of $E(t)$. This is chosen so that the probability per time to count a single photon in this mode is $|E(t)|^2$ for a unit efficiency detector.   A simple way to model this is to include a classical driving term  in the interaction Hamiltonian as $H_d=\epsilon(t)(b+b^\dagger)$, with $\epsilon(t)=\sqrt{\kappa_b}E(t)$ The input field to mode $a$ is not a vacuum state, but non stationary pulsed single photon states. In this case an  equivalent formulation in terms of the quantum stochastic differential equations enables us to refer directly to the fields external to the cavity.  These take the form\cite{Walls-Milb,Gard_Zol}
\begin{eqnarray}
\label{qsde}
\frac{d a}{dt} & = & -i\delta_a a-i\chi \hat{n}_b(t) a-\frac{\kappa_a}{2} a+\sqrt{\kappa_a}a_i(t)\\
\frac{db}{dt}  & = & -i\epsilon(t)-i\chi \hat{n}_a(t) b-\frac{\kappa_b}{2} b+\sqrt{\kappa_b}b_i(t)
\label{qsde}
\end{eqnarray}
where the number operators are  $\hat{n}_a(t)\equiv a^\dagger(t)a(t),\ \ \ \hat{n}_b\equiv b^\dagger (t) b(t)$, and $\chi$ is proportional to the third order optical nonlinearity, while $\delta_a=\omega_a-\Omega_a$, where $\omega_a$ is the cavity frequency for mode-a, and with $\kappa_a,\kappa_n$ the photon number decay rates for modes $a$ and $b$ respectively. 
In this formulation the quantum noise terms in fact represent the input modes to the cavity. In this case, the noise term $b_i(t)$ is vacuum noise, however the noise for $a_i(t)$ requires careful treatment. We can also define output modes, in the interaction picture, 
\begin{eqnarray}
a_o(t) & = & \frac{1}{\sqrt{2\pi}}\int_{-\infty}^\infty d\omega a_o(\omega) e^{-i\omega t}\\
b_o(t) & = & \frac{1}{\sqrt{2\pi}}\int_{-\infty}^\infty d\omega b_o(\omega) e^{-i\omega t}
\end{eqnarray}
which represent the quantum noise terms for the time reversed versions of the quantum stochastic differential equations\cite{Gard_Zol}. The input and output operators are related through the cavity modes
\begin{equation}
a_o(t)  =  \sqrt{\kappa_a} a(t)-a_i(t)
\label{a-out}
\end{equation}
\begin{equation}
b_o(t)  =  \sqrt{\kappa_b} b(t)-b_i(t)
\label{b-out}
\end{equation}

The standard approach to solving these equations is to take the Fourier transform and neglect the initial transients to obtain algebraic equations for $a(\omega), a_i(\omega)$ etc. If the equations are linear, the output statistics can then be directly determined from the input statistics. However here the equations are non linear and this approach will not work. We will follow a different approach inspired by the quantum trajectory formulation in terms of realisations of the quantum stochastic differential equations.  

 Our approach is based on replacing $\hat{n}_a(t)$ by a classical stochastic process $n_a(t)$. This is a stochastic binary variable. The transition rate for the process $n_a=0\rightarrow n_a=1$ is  the {\em absorption} rate for the single photon in the input state to mode-$a$. As discussed below,  this is given by Eq.(\ref{abs-rate}).  The transition rate for the process $n_a=1\rightarrow n_a=0$ is  the {\em emission} rate for the single photon in the input state to mode-$a$. In effect this way of thinking treats the cavity for mode-a {\em as if} it was a photon detector. Of course  no real record is available of this detection event, but as it is an irreversible process we can still think of it as a detection event in principle. 

We first calculate the absorption rate, that is, the probability per unit time that the cavity makes a transition from the vacuum state to a one photon state $|0\rangle_a\rightarrow |1\rangle_a$.  The stochastic differential equation for mode-a is given by Eq(5). We will now make an approximation $\hat{n}_b(t)\approx n_b(t)\sim |\epsilon(t)|^2$. This means we are ignoring the effect of quantum fluctuations of the coherent field, mode-b, on the single photon field. We will address the validity of this approximation in section \ref{cas}  where we show that it is justified so long as $\chi$ is the smallest frequency in the problem. The stochastic differential equation for mode-a is then
\begin{equation}
\frac{da}{dt}=-i\chi n_b(t)a-\frac{\kappa_a}{2}a+\sqrt{\kappa_a}a_i(t)
\end{equation}
where $n_b(t)=\kappa_b|E(t)|^2$ is a known function.  The formal solution is
\begin{equation}
a(t)=a(0)e^{-\kappa_a t/2-i\xi(t)}+\sqrt{\kappa_a}e^{-\kappa_a t/2-i\xi(t)}\int_0^t dt' a_i(t') e^{\kappa_a t'/2+i\xi(t')}
\end{equation}
with 
\begin{equation}
\xi(t)=\chi \int_0^t dt'  n_b(t')+\delta_a t
\end{equation}
Given that the cavity field at time $t=0$ is not expected to be correlated with the input field at time $t>0$, we see that
\begin{eqnarray}
\label{modea_num}
\frac{d\langle \hat{n}_a\rangle}{dt} & = &  -\kappa_a \langle \hat{n}_a(t)\rangle\\\nonumber
 & & +\kappa_a e^{-\kappa_a t/2+i\xi(t)}\int_0^t dt' \langle a_i^\dagger(t')a_i(t)\rangle e^{\kappa_a t'/2-i\xi(t')} \\\nonumber
 	& & 	+\kappa_a e^{-\kappa_a t/2-i\xi(t)}\int_0^t dt' \langle a_i^\dagger(t)a_i(t')\rangle e^{\kappa_a t'/2+i\xi(t')}
\end{eqnarray}
The two time correlation function for the input field can be explicitly evaluated given the input state, Eq.(\ref{single-photon});
\begin{equation}
 \langle a_i^\dagger(t_1)a_i(t_2) \rangle=\nu^*(t_1)\nu(t_2)
 \end{equation}
The first term in Eq.(\ref{modea_num}) corresponds to emission of photons from the cavity at the Poisson rate $\kappa_a$. The second term is thus the absorption rate, or probability per unit time to absorb the single photon into the cavity, which we write as $p_{abs}(t)$, so that
\begin{equation}
p_{abs}(t)  =  \kappa_a e^{-\kappa_a t/2+i\xi(t)}\int_0^t dt' \nu^*(t')\nu(t) e^{\kappa_a t'/2-i\xi(t')}+{\rm c.c}
\end{equation}		
which is completely determined by the decay rate for mode-a, and the input fields for modes a and b. In the special case that $\chi=0$ and $\nu(t)$ is real this reduces to 
\begin{equation}
p_{abs}(t)=2\kappa_a e^{-\kappa_a t/2}\int_0^t dt' \nu(t')\nu(t) e^{\kappa_a t'/2} 
\label{abs-rate}
\end{equation}
If mode-b is driven by a continuous field, $n_b(t)$ can be replaced by its steady state value, $n_b$. If we then detune the mode-a cavity so that $\delta_a=-\chi n_b$ we can then set $\xi(t)=0$ and use the result in Eq.(\ref{abs-rate}) for the absorption rate in the presence of the nonlinearity. This means that the mode-a cavity is on resonance even when the Kerr nonlinearity due to  mode-b is taken into account. 

As an example we consider 
\begin{equation}
\nu(t)=\left \{\begin{array}{cc}
				\sqrt{\gamma}e^{-\gamma(t-t_0)/2} & t\geq t_0\\
				0 & t< t_0
				\end{array}\right .
\label{cavity-source}
\end{equation}
which corresponds to a single photon being emitted  from a cavity, with damping constant $\gamma$, when prepared in a single photon state at time $t_0$, which we now set to zero, $t_0=0$.  The absorption rate is then given by
\begin{equation}
p_{abs}(t)=\frac{4\kappa_a}{\gamma-\kappa_a}\left (e^{-(\kappa_a-\gamma)t/2}-1\right )(\gamma e^{-\gamma t})
\label{qsde-abs}
\end{equation}
For both cases,  $\kappa_a>\gamma$ and $\gamma >\kappa_a$ it is positive.

If we substitute the absorption rate into the equation for the mean photon number in mode-a, Eq.(\ref{modea_num}), we can solve it directly to give
\begin{equation}
\langle n_a(t)\rangle =\frac{4\gamma\kappa_a}{(\gamma-\kappa_a)^2}e^{-\kappa_a t}\left (1- e^{-(\gamma-\kappa_a)t/2}\right )^2
\label{num-a}
\end{equation}
This is the ensemble average photon number in mode-a, averaged over all absorption and emission events.  

\begin{figure}[!hbt]
\begin{center}
\includegraphics[scale=0.6]{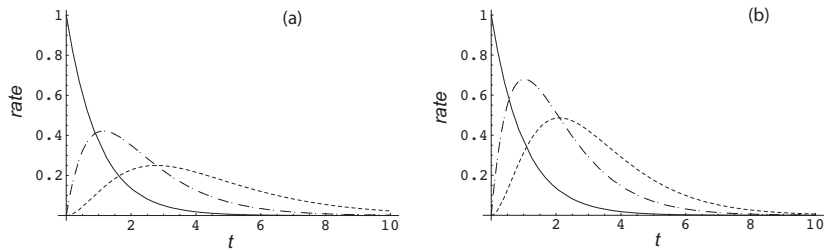}
\end{center}
\caption{The plot of the source emission rate (solid), the absorption rate (dash-dot) and  emission rate (dot) of the  mode-a cavity for, (a) $\kappa_a=0.5$, (b) $\kappa_a=0.9$. In both cases $\gamma=1.0$  }
\label{fig-2}
\end{figure}

To assist with the interpretation of the various probabilities we first note that in our model we only calculate time dependent probabilities and rates; there is no spatial dependence. This is the point of view taken in the input-output formalism that has proved to be so effective in quantum optics.  There are three mutually exclusive ways in which a photon can be detected between times $t$ and $t+dt$:  (i) it can be detected in the source with probability $Pr_s(t)$,  (ii) it can be detected in the cavity with probability $Pr_c(t)$ and (iii) it can be detected outside the source {\em and} the cavity with probability $Pr_o(t)=1-(Pr_s(t)+Pr_c(t))$. In our model it is easy to see that 
\begin{eqnarray}
Pr_s(t) & = & e^{-\gamma t}\\
Pr_c(t) & = & \langle \hat{n}_a(t)\rangle 
\end{eqnarray}
where the last equation follows from the fact that there can only ever be at most one photon in the cavity. It then follows that the rate of detection of photons outside the cavity must be
\begin{equation} 
p_o(t)=\frac{dPr_o(t)}{dt}=\gamma e^{-\gamma t}-\frac{d\langle\hat{n}_a\rangle}{dt}
\end{equation}
Using Eq. (\ref{modea_num}) together with the interpretation implicit in Eq (\ref{abs-rate}) we then see that 
\begin{equation} 
p_o(t)=|\nu(t)|^2-p_{abs}(t)+\kappa_a\langle\hat{n}_a(t)\rangle
\label{rate-out}
\end{equation}
On the other hand, the input-output formulation indicates that the rate of detection for photons after an interaction with the cavity is 
\begin{equation}
p_o(t)= \langle n_o(t)\rangle =\langle a_o^\dagger(t)a_o(t)\rangle
\end{equation}
If we use the input/output relation in Eq.(\ref{a-out}) we find that 
\begin{equation}
\langle n_o(t)\rangle =|\nu(t)|^2-p_{abs}(t)+\kappa_a\langle n_a(t)\rangle 
\end{equation}
which agrees with Eq.(\ref{rate-out}).  This simply says that the rate of counting photons after the cavity is the  rate at which photons are generated, minus the rate they are absorbed by the cavity plus the rate at which they are emitted from the cavity. In figure \ref{fig-2} we plot the source emission rate, the absorption rate and  emission rate of the  mode-a cavity for two values of $\kappa_a$. In both cases $\gamma=1.0$. 

We return now to the calculation of the phase shift on mode-b produced by the single mutual Kerr interaction with mode-a, driven by a single photon source.  In a quantum trajectory picture, the number of photons in mode-a is a classical stochastic process $n_a(t)$ which takes the values zero or unity.  The nonlinear quantum stochastic differential equation, Eq.(\ref{qsde}),  can now be replaced by
\begin{equation}
\frac{db}{dt}   =  -i\epsilon(t)-i\chi n_a(t) b-\frac{\kappa_b}{2} b+\sqrt{\kappa}b_i(t)
\label{qt-qsde}
\end{equation} 
We consider a continuous wave for mode-b and set the initial condition long before the single photon pulse reaches the cavity, so that mode-$b$ dynamics is given by Eq.(\ref{qt-qsde}) with $n_a=0$.  In that case,  the mode-$b$ will have reached  a steady state which is in fact a coherent state with amplitude $
\beta_\infty=\frac{-2i\epsilon}{\kappa_b}$. 

The single photon input to mode-$a$ will have entered the cavity at some random time $t_A$ and $n_a(t)=1$ for $t>t_A$.  At some Poisson distributed time, $t_E$,  the photon will be emitted from the cavity, at rate $\kappa_a$, so for $t>t_E$ we have $n_a(t)=0$.  In that case, $T=t_E-t_A\equiv T$, is a random variable corresponding to the dwell time of the mode-$a$ photon in the cavity. Over this time interval the state of mode-$b$ evolves as a coherent state with amplitude,
\begin{equation}
\beta(T)=\frac{-2i\epsilon}{\kappa_b+2i\chi}+\frac{4\epsilon\chi}{\kappa_b(\kappa_b+2i\chi)}e^{-(\kappa_b/2+i\chi)T}
\end{equation} 
where $T$ is a classical random variable. We can determine the change in the phase of the field from the steady state value, $\Delta\beta(T)=\beta(T)-\beta_\infty$ as 
\begin{equation}
\Delta\beta(T)=\frac{4\epsilon\chi}{\kappa_b(\kappa_b+2i\chi)}\left (e^{-(\kappa_b/2+i\chi)T}-1\right )
\label{delta-amp}
\end{equation}
We expect that in typical situations that $\kappa_b>> \chi$ so that we can approximate
\begin{equation}
\Delta\beta(T)=\frac{4\epsilon\chi}{\kappa_b^2}\left (e^{-\kappa_bT/2}-1\right )
\label{amp-change}
\end{equation}
This is the {\em conditional} change in the amplitude of the probe field from steady state, conditioned on a photon having entered mode-a.

\begin{figure}[!hbt]
\begin{center}
\centering{\includegraphics[scale=0.6]{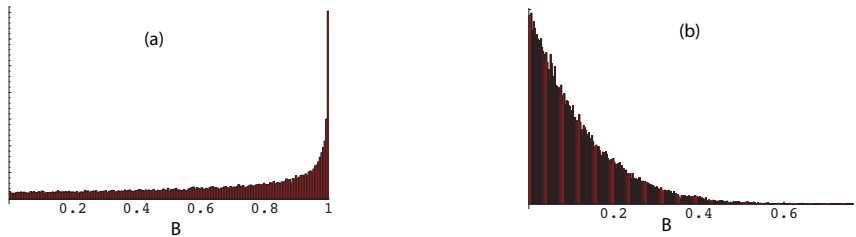}}
\end{center}
\caption{A histogram of the changes in the amplitude of mode-b from the steady state value for many realisations of the time, $T$, over which the single photon resides in the cavity for, $4\epsilon\chi/\kappa_b^2=1$  and (a)$\kappa_b/\kappa_a=4.0$, (b)$\kappa_b/\kappa_a=0.25$. The number of trials is $50,000$. }
\label{simul}
\end{figure}

The statistics of $T$ is determined by the absorption and emission of the photon by the cavity. Recalling that we are in a 'quantum trajectory' picture in which $n_a(t)$ is zero or unity, the conditional emission probability {\em given} that $n_a(t)=1$ is a Poisson process with rate $\kappa_a$. The probability distribution for $T$ is then given by
\begin{equation}
p(T) =\kappa_a e^{-\kappa_a T}
\end{equation}
The average dwell time, $T$, is then simply $(\kappa_a)^{-1}$. The change in the amplitude, Eq.(\ref{amp-change}) will thus be greatest when $\kappa_a/\kappa_b<<1$. Furthermore the fluctuations in amplitude change will also be small. We can test 
this intuition by simulating the Poisson process for $T$.  In figure \ref{simul} we show histograms of $|\Delta\beta(T)|$ given in Eq.(\ref{amp-change}), with $\epsilon\chi/\kappa_b^2=1$. In figure \ref{simul}(a) we show an example in which $\kappa_b/\kappa_a< 1$. The wide variance in the amplitude shift is evident, whereas in figure \ref{simul}(b), for which $\kappa_b/\kappa_a> 1$ and $\chi/\kappa_b$,  the change in amplitude is peaked at $\frac{4\epsilon\chi}{\kappa_b^2}$.
The fluctuations in the displacement $\Delta\beta(T)$ result from the photon leaving mode-a before mode-b has a significant chance to respond. 

We can now compute the ensemble average of the {\em conditional} amplitude change,  $\Delta\beta(T)$, over the distribution of waiting times, $T$. The mean is given by
\begin{equation}
{\cal E}[\Delta\beta(T)]= -B\left (\frac{\kappa_b}{\kappa_b+2\kappa_a}\right )
\label{cond-mean}
\end{equation}
while the variance is 
\begin{equation}
{\cal E}[\Delta\beta(T)^2]-{\cal E}[\Delta\beta(T)]^2 =B^2\frac{2\kappa_a\kappa_b}{(\kappa_b+2\kappa_a)^2} 
\end{equation}
where 
$
B=\frac{4\chi\epsilon}{\kappa_b^2}
$.
In the limit $\kappa_b>>\kappa_a$ we find that the conditional mean is $-B$ and the variance goes to zero. This result is a conditional mean conditioned on the fact that the single photon does indeed enter the cavity. We now wish to compute the {\em unconditional} average of the change in the amplitude for the probe, at time $t$, averaged over all emission and absorption events for mode-a. We will do this using the cascaded systems formalism of quantum optics.

\section{Cascaded systems approach.}
\label{cas}
We can also describe the system in terms of an ensemble average over the random absorption and emission times of the single photon using the cascaded systems method of Carmichael and also Gardiner \cite{Car,Gar}. One cavity, labeled by annihilation operator, $c$, is called the {\em source}, the output from which is coupled uni-directionally into the second cavity for the two modes $a$ and $b$ coupled by a cross Kerr interaction as before.  The system is described, in the interaction picture, by the master equation 
\begin{eqnarray}
\frac{d\rho}{dt} & = & -i\delta_a[a^\dagger a,\rho]-i\epsilon[b+b^\dagger,\rho]-i\chi[a^\dagger a b^\dagger b,\rho]\\\nonumber
& & + \gamma{\cal D}[c]\rho+\kappa_a{\cal D}[a]\rho+\kappa_b{\cal D}[b]\rho+\sqrt{\gamma \kappa_a}\left ([c\rho,a^\dagger]+[a,\rho c^\dagger]\right )
\label{cas-me}
\end{eqnarray}
We assume that the source cavity starts at time $t=0$ in a one photon state $|1\rangle_c$ and first consider the situation with $\chi=0$, i.e we only consider mode-a.   The equation of motion for the number in the driven cavity is then
\begin{equation}
\frac{d\langle \hat{n}_a\rangle}{dt}=-\kappa_a\langle \hat{n}_a\rangle-\sqrt{\gamma\kappa_a}(\langle a^\dagger c\rangle+\langle c^\dagger a\rangle)
\label{cas-num}
\end{equation}
The second term is the absorption probability in analogy to the second term in Eq(\ref{modea_num}). The equation of motion for $\langle c^\dagger a\rangle$ can be solved directly and substituted into Eq.(\ref{cas-num}) to obtain
\begin{equation}
\frac{d\langle \hat{n}_a\rangle}{dt}=-\kappa_a\langle n_a\rangle+\frac{4\gamma\kappa_a}{(\gamma-\kappa_a)}e^{-(\gamma+\kappa_a)t/2}\left (1-e^{-(\gamma-\kappa_a)t/2)}\right )
\label{cas-num2}
\end{equation}
This implies the absorption probability per unit time for a single photon to be absorbed from the source cavity is the second term and this  agrees with the result in  Eq.(\ref{qsde-abs}).   We can solve Eq.(\ref{cas-num2}) directly to obtain the result quoted in Eq.(\ref{num-a}). 

We assume that mode-b starts at $t=0$ in the steady state coherent state $|\beta_\infty\rangle$ while mode-c starts in a single photon state $|1\rangle_c$ while mode-a starts in the vacuum state. We now make a canonical transformation to a {\em displacement picture} $\bar{\rho}(t)=D(\beta_\infty)\rho(t)D^\dagger(\beta_\infty)$. This shifts mode-b to a new equilibrium position so that at $t=0$ mode-b starts in the vacuum state as well as mode-a. The master equation in the displaced picture becomes,
\begin{eqnarray}
\frac{d\rho}{dt}& = & -i\chi[a^\dagger a (\beta_\infty b^\dagger+\beta^*_\infty b,\rho]-i\chi[a^\dagger a b^\dagger b,\rho]\\\nonumber
& & + \gamma{\cal D}[c]\rho+\kappa_a{\cal D}[a]\rho+\kappa_b{\cal D}[b]\rho+\sqrt{\gamma \kappa_a}\left ([c\rho,a^\dagger]+[a,\rho c^\dagger]\right )
\label{cas-me}
\end{eqnarray}
We also assume that the detuning $\delta_a$ has been chosen to cancel any systematic detuning due to the back action of the probe field. 
We remind the reader at this point that using the master equation in Eq.(\ref{cas-me}) describes the unconditional dynamics of the system, averaged over all possible emission and absorption events from each cavity.  We thus do not expect to be able to reproduce the conditional phase shifts discussed in the previous section. However we can calculate ensemble average phase shifts on the mode-b using this master equation. 

The equations of motion for the mean amplitude in mode-b is
\begin{equation}
\frac{d\langle b\rangle_d}{dt} = -i\chi\beta_\infty \langle \hat{n}_a\rangle_d-\frac{\kappa_b}{2}\langle b\rangle_d-i\chi\langle \hat{n}_a b\rangle_d
\label{cas-amp}
\end{equation} 
where the $d$ subscript reminds us that we are in a displaced frame in which $\beta_\infty$ has been subtracted. 
The equation of motion for $\langle \hat{n}_a\rangle$ is still given by Eq.(\ref{cas-num}), however there is now an implicit $\chi$ dependence, as the equation of motion for $\langle c^\dagger a\rangle$ depends on $\chi$.  If we neglect this dependence and approximate $\langle \hat{n}_a\rangle$ in Eq.(\ref{cas-amp}) with the solution in Eq.(\ref{num-a})  we are in effect seeking a solution to linear order in $\chi$.  We also assume that $|\beta_\infty|>>1$ and neglect the final term in Eq.(\ref{cas-amp}). This is justified as the state of mode-a is at most one photon away from the vacuum while to lowest order in $\chi$, mode-b is near the vacuum state in the displacement picture. Neglecting this term means we are ignoring correlation between mode-a and mode-b and is consistent with the assumption that mode-b remans in a coherent state as we assumed in the section \ref{SDE}.  The solution for $\langle b\rangle $ in the displacement picture is then easily obtained.  However in section \ref{SDE} we saw that best performance is likely to require that mode-b is slaved to mode-a, that is to say $\kappa_b>>\kappa_a$. In that case we see that the change in amplitude of mode-b in the ensemble average is, to linear order in $\chi$,  
\begin{equation}
\Delta\beta(t)=\langle b\rangle_d=-\frac{4\epsilon\chi}{\kappa_b^2}\langle \hat{n}_a(t)\rangle 
\end{equation}
Of course this eventually decays to zero as the photon in mode-a eventually leaves the cavity and mode-b simply relaxes back to the steady state coherent state of amplitude $\beta_\infty$.  In contrast the condition mean in Eq.(\ref{cond-mean}) assumes that the photon has entered the cavity but has {\em not} left the cavity.

\section{Discussion and conclusion.}
\begin{figure}[!hbt]
\begin{center}
\centering{\includegraphics[scale=0.6]{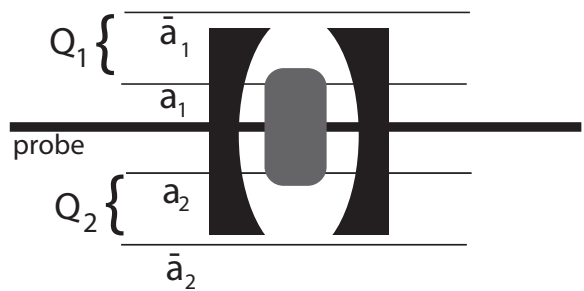}}
\end{center}
\caption{A model of a two-qubit parity gate for dual rail code with single photon states. The qubits are labeled as $Q_k$. The probe field is a strong coherent state. The box represents a cavity resonant with both the probe and the qubit modes. One rail of each qubit does not pass though the cavity.   }
\label{photon-parity}
\end{figure}

One of the important applications of the model of this paper is to enable a two-qubit parity gate based on an intracavity weak nonlinearity, (see figure \ref{photon-parity} ). The qubits are encoded as dual rail single photon states, for example, each rail could correspond to orthogonal polarisation states of a single spatial mode. In figure  \ref{photon-parity} we distinguish the modes of each rail using an overbar notation, $a_k,\bar{a}_k$, where $k=1,2$ for each qubit. Only one rail of each qubit is directed towards the cavity inside of which we model the interaction as 
\begin{equation}
H=\hbar\chi(a_1^\dagger a_1-a_2^\dagger a_2)b^\dagger b
\end{equation}
We are not suggesting this as a realistic description of a mutual Kerr nonlinearity but rather as a minimal model for a two qubit parity gate for dual rail single photonic qubits.  Our objective here is to assess the effect on the gate of the random interactions times of each photon with the coherent field in the cavity. The parity gate works as follows. If the number of photons in each mode are the same, that is to say both $0$ or both $1$, there can be no phase shift on the coherent probe field. If however the photon number in each mode is different, there is an equal and opposite phase shift for the two cases $n_1=1,n_2=0$ and $n_1=0,n_2=1$. This assumes that each photon interacts with the probe field for the same time. However  as we saw in the single mode case, the absorption and emission of photons from the cavity are stochastic events. Furthermore in this model the stochastic events for each mode are independent (if we ignore the back action of the probe field on the single photon modes).   

Using the results in the previous section, the probe field is injected as a coherent state with amplitude $\epsilon(t)/\sqrt{\kappa_b}$, in which case it will remain a coherent state with amplitude $\beta(t)$ given by
\begin{equation}
 \frac{d\beta}{dt}= -i\epsilon(t)-i\chi(n_1(t)-n_2(t))\beta-\frac{\kappa_b}{2}\beta
 \end{equation}
 where $n_1(t), n_2(t)$ are well approximated by independent stochastic variables that take values $0,1$ and correspond to the number of photons in cavity modes $a_1,a_2$ respectively.  We again assume that $\epsilon(t)=\epsilon$ is a constant and that the probe field has relaxed to a steady state long before the qubit photons enter the cavity. We also assume that the qubit photons are identical synchronous single photon pulsed states given by Eq. \ref{single-photon}.  As before we expect the best performance to occur when the probe field is heavily damped so that it is slaved to the qubit modes. In the typical case for which $\kappa_b>>\chi$ we then find that the change in the amplitude of mode-b from the steady state value is given by 
 \begin{equation}
 \Delta\beta(t)=-\frac{4\chi\epsilon}{\kappa_b^2}(n_1(t)-n_2(t))
 \end{equation}
For a perfect parity gate we would like this amplitude change to be zero for the two cases in which both modes had either zero or one photon. However it is now clear that this can only be true on average. From shot to shot, we are not guaranteed that the stochastic variables both take the same values even when both qubits have zero or one photon in both of the cavity rails and so, from shot to shot, the amplitude change will fluctuate around the ideal case of zero. Likewise for the case in which the number of photons in each of the cavity rails is different.  In that case would like there to be a significant equal and opposite amplitude change, but now there will be fluctuations superimposed.  A detailed error analysis of quantum circuits based on the effect of these fluctuations will be the subject of  a future paper.

In this paper we have presented two approaches to nonlinear intracavity systems driven by single photon states. One approach is based on a generalization of the quantum Langevin approach to the case where the input noise describes single photon excitations of the extra-cavity field. The second approach is based on the cascaded system master equation of Carmichael\cite{Car} and Gardiner\cite{Gar}. In both cases we see intrinsic noise arising from the random absorption and emission times of the photon. These processes are however modeled as coherent interactions between the fields inside and outside the cavity. As a simple application of the formalism we discussed a two-photon parity gate based on weak nonlinearities, however many more examples could be given. Of particular interest is the case of single photons interacting with cavities containing single atomic systems. This situation arises very naturally in the context of quantum repeaters\cite{repeaters}, a key component of a quantum internet based on single photonics\cite{Kimble}.  Given the effort going into making controllable single photon sources we expect that single photon quantum optical systems of the kind discussed in this paper will play an increasingly important part in the future of the field.   

\section*{Acknowledgements}
This work was supported in part by MEXT, NICT and JSPS in Japan and the EU projects HIP and QAP, and also by the Australian Research Council Centre for Quantum Computer Technology.

\end{document}